\documentclass[12pt]{article}
\usepackage[english]{babel}
\usepackage{epsf}
\usepackage{pazh}
\usepackage{amsmath}	% Advanced maths commands
\usepackage{amssymb}	% Extra maths symbols
\newcommand{\ergs}{\,erg\,s$^{-1}$}
\newcommand{\kms}{\,km\,s$^{-1}$}

\newcommand{\gcm}{\,g\,cm$^{-3}$}

\begin{document}
\title{\bf $^7$Be abundance in nova V5668 Sgr doesn't contradict theory
 }
\author{\bf \hspace{-1.3cm}\copyright\, 2020 \\
N. N. Chugai$^{1}$
and
A. D. Kudryashov$^{2}$
}
\affil{
$^{1}$Institute of Astronomy, Russian Academy of Sciences, Moscow\\ 
}
\affil{$^{2}$Russian Institute of scientific and technical Information, 
Russian Academy of Sciences, Moscow\\ 
}

\vspace{2mm}
%\received{\today}

\sloppypar 
\vspace{2mm}
\noindent


\vfill
\noindent\rule{8cm}{1pt}\\
{$^*$ email: $<$nchugai@inasan.ru$>$}

\clearpage

\begin{abstract}
\centerline{\bf Abstract}
Resonance lines of $^7$Be are detected currently in five novae. 
Available abundances for this isotope estimated from equivalent 
 widths of $^7$Be\,II and Ca\,II lines are significantly higher  
 compared to predictions of models for the thermonuclear flash.
In attempt to pinpoint the reason for this disparity we explore the possibility 
 for the higher $^7$Be yield via computing kinetics of the thermonuclear burning in
 the framework of two-zone model and find that even for a favorable choice of 
 parameters $^7$Be mass fraction does not exceed $3\cdot10^{-5}$. This is consistent 
 with known theoretical results and leaves the disparity between the theory and 
 observations unresoved.
We find that the contradiction is caused by the assumption that the ionization 
  fraction of Be\,II/Be is equal to that of Ca\,II/Ca, which has been adopted 
  formerly 
  in order to estimate the $^7$Be abundance.
In the case of nova V5668 Sgr the ionization fraction of Be\,II/Be turns out 
  to be at least by a factor of $\sim 10$ higher compared to Ca\,II/Ca due to 
  difference of ionization potentials.
Our new estimate of the $^7$Be mass fraction for nova V5668 Sgr does not contradict 
the theory.
The calculated flux of 478 keV gamma-quanta from the $^7$Be decay is consistent 
with the upper limit according to {\em INTEGRAL} observations.
\end{abstract}

\section{Introduction}

The nova outburst is caused by a flash of the thermonuclear burning of 
  the hydrogen-rich matter accumulated on the white dwarf via the accretion in 
  a binary system. 
The outburst results in the synthesis of large variety of radioactive isotopes including 
  $^7$Be. 
Cameron (1955) proposed that stars could be sources of $^7$Li in the Galaxy owing 
  to the reaction $^3\mbox{He}(\alpha,\gamma)^7\mbox{Be}$ followed by 
  decay via $e$-capture $^7$Be + e$^{-}$ $\rightarrow$ $^7$Li + $\nu + \gamma$ 
  ($t_{1/2} = 53.12$, Firestone et al. 1999).
 The $^7$Li production by novae via the Cameron process first considered by 
   Starrfield et al. (1978) in a framework of the hydrodynamics with the kinetics 
   of nuclear reactions. 
 Hydrodynamic models of thermonuclear flash on CO white dwarf (Starrfield et al. 2019, 
  Hernanz et al. 1996, Jos{\'e} and Hernanz 1998, Denissenkov et al. 2014) 
  predict $^7$Be mass fraction in novae ejecta 
  X$(^7\mbox{Be}) \sim 10^{-6} - 2\cdot10^{-5}$.

Recently $^7$Be was detected in five novae via identification of resonance doublet 
  of $^7$Be 3130.4219, 3131.0667\,\AA\ (Tajitsu et al. 2015, Molaro et al. 2016, 
  Selvelli et al. 2018, Izzo et al. 2018).
The $^7$Be abundance estimates in these papers significantly (upto 1 dex) exceed 
  theoretical predictions of the nucleosynthesis during novae outbursts.
This controversy has not been debatable as yet, despite the issue is crucial both for the 
 test of nucleosynthesis in novae and for the assessment of the novae contribution in 
 galactic lithium.
Meanwhile two reasons for this disparity are conceivable: either a model drawbacks, 
  e.g., due to simplified treatments of convective mixing during the accretion 
 and the nuclear flash, or the incorrect interpretation of absorption 
 lines of $^7$Be in terms of the abundance.
In this regard some doubts immediately can arise concerning the 
  equality of ionization fractions of Be\,II/Be and Ca\,II/Ca that has been assumed 
  for the estimation of the $^7$Be abundance. 

In the present paper we pursue a goal to disantangle controversy between 
  theoretical predictions and observational estimates of the $^7$Be abundance.
We explore two options.
First, we address the issue, what is the maximal $^7$Be abundance expected 
  in novae shell as a result of the thermonuclear flash.
To this end we use a two-zone model of the thermonuclear burnig that permits us 
  to explore a large volume of parameter space.
The second approach is aimed at the determination of the $^7$Be abundance 
  from absorption lines of the $^7$Be resonance doublet.
We concentrate on the nova V5668 Sgr discovered at 2015 March 15 (Seach 2015).
The high resolution spectrum with high S/N ratio on day 58 after the maximum  
   shows both lines of the resonance doublet $^7$Be\,II 3130.42, 3131.07\,\AA\ 
   at the radial velocity of -1175\kms\ (Molaro et al. 2016).
Observational data in the optical, infrared, ultraviolet, and X-ray band provide us 
  with the estimate of the bolometric luminosity at the stage of interest 
  (Gherz et al. 2018), which reduces unsertainties of our results.
Remarkably, this nova was observed in the X-ray band by spectrograph SPI onboard 
 {\em INTEGRAL} with the recovered upper limit of the flux in the 478 keV line 
 (Siegert et al. 2018). 
  
Following to claimed goals we describe two-zone model of the thermonuclear flash 
  (Section \ref{sec:nuc}) and compute the shell composion including the $^{7}$Be 
  abundance for various choice of parameters.
In section  \ref{sec:abun} we present a new approach for the estimation of the 
  $^{7}$Be abundance in nova V5668 Sgr with the computed ionization fractions 
  of Be\,II and Ca\,II.
Below we distinguish between ages counted from the optical maximum and from 
 the thermonuclear flash that presumably took place 7 days prior the optical 
 maximum (Siegert et al. 2018); the latter time is used for the determination of the 
  $^{7}$Be decayed fraction.

\section{$^7$Be synthesis in novae}
\label{sec:nuc}

\subsection{Model}

The two-zone model applied here is similar to that of Boffin et al. (1993).
It consists of the central high-temperature zone and external low-temperature 
  zone.
The mass ratio of the central-to-external zone is 1/10, close to the value 1/9 
  adopted by Boffin et al. (1993).
We find that resulting $^7$Be abundance almost insensitive to this parameter.
The adopted evolution of the temperature and density in the central zone obeys 
  adiabatic law ($\rho\propto T^3$) with the exponential temporal behavior 
  $\rho = \rho_0\exp{(-t/t_e)}$ ($t_e$ is called burning time).
The temperature and density of the external zone is determined via fixed 
  ratio of temperatures of the central and external zones ($T_2/T_1$).
Maximal values $\rho_0$ and $T_0$ vary in the ranges of $(10^2-10^4)$\gcm\ and 
  $(1-5)\cdot10^8$\,K, while the burning time $t_e$ varies in the range of $100-300$\,s.
An effect of the convective mixing between zones is described in terms of a constant 
 rate of the mass exchange with the mixing time ($t_c$) for the external zone adopted  
 to be in the range of $10^2-10^4$\,s.
The fraction of the admixed white dwarf matter ($q$) varies in the range of 
  $0-0.5$.
 It should be emphasised that the energy considerations require $q > 0.1$ 
  (Starrfield et al. 1978, 2019).
Adopted ranges of parameters correspond to physical conditions during thermonuclear flash 
  in hydrodynamical models (e.g., Denissenkov et al. 2014).
 
In each zone we solve the system of kinetic equations for mole fraction 
   $Y_i=X_i/A_i$, where $X_i$ and $A_i$ are the mass fraction and atomic weight of 
  an $i$-th nucleus.
We take into account 282 nuclei between $^1$H and $^{57}$Cr that are involved in 
  2011 reactions of nuclear burning.
The composition of accreted matter is solar.
The composition of the CO dwarf is the same as adopted earlier (Kudryashov et al. 2000), 
 while for the ONeMg white dwarf the composition is taken according to Ritossa et al. (1996).
Computations show that the $^{7}$Be abundance in our models does not depend on the 
  type of the white dwarf.
Nuclear reaction rates are taken from databases NACRE (Angulo et al. 1999) and 
  JINA REACLIB (Cyburt et al. 2010).

\subsection{Modelling results}

$^{7}$Be abundance at the end of active burning stage, i.e., when the 
  initial temperature decreased by a factor of 10, is shown in Fig. 1 as a function 
  of $q$.
Panels {\em a, b, c, d} one-by-one show effects of initial temperature of the central 
  zone, mixing time for the external zone, initial density of the central zone, and 
  the temperature ratio for the central-to-external zone.

The $^{7}$Be abundance weakly depends on the initial temperature (Fig. 1, panel {\em a})
  and on the mixing time (Fig. 1, panel {\em b}).
The four-fold density decrease results in the variation of the  $^{7}$Be abundance
  in the range of $10^{-6} - 8\cdot10^{-6}$ (Fig. 1, panel {\em c}).
Even more pronounced is the effect of the temperature ratio (Fig. 1, panel {\em d}). 
For the higher $T_2/T_1$ ratio the $^{7}$Be abundance gets higher and approaches to 
  X($^7$Be) $\approx 3\times10^{-5}$ for $q < 0.1$.
For $T_2/T_1 > 3$ the  $^{7}$Be abundance does not change (Kudryashov 2019, Fig.2). 
We therefore consider X($^7$Be) = $3\times10^{-5}$ to be the theoretical upper 
  limit of $^{7}$Be abundance in nova ejecta.
For the optimal value of $q\approx 0.25$ (Starrfield et al. 2019) the maximal
 $^{7}$Be abundance in our models is $2.2\cdot10^{-5}$ (Fig. 1, panel {\em d}). 
Remarkably that this value is almost coinsides with the maximal value of 
  X($^7$Be) = $2\cdot10^{-5}$ for CO white dwarfs with masses 1.15-1.35\msun\ 
  in recent hydrodynamic models (Starrfield et al. 2019).

\section{$^7$Be abundance in nova V5668 Sgr}
\label{sec:abun}

Using the ratio of equivalent width of $^7$Be\,II and Ca\,II 3933\,\AA\ lines 
  in nova V5668 Sgr on day 58 Molaro et al. (2016) finds  the  number ratio 
  $N(^7\mbox{Be})/N(\mbox{Ca}) \approx 53-69$.
This implies the mass fraction of $^7$Be\,II  of $\sim 9\cdot10^{-5}$, i.e., 
  three times larger than the upper limit found above. 
The $^7$Be abundance in nova V5668 Sgr is found based upon the assumption that 
 the ionization fraction of $^7$Be\,II/$^7$Be and Ca\,II/Ca are equal.
This assumption however is doubtful because ionization potential of 
  Be\,II and Ca\,II as well as their ions of higher ionization states are essentially 
  different. 
Below we estimate $^7$B abundance taking into account ionization fractions of  
 of Be\,II and Ca\,II.  

\subsection{$^7$Be column density}
\label{sec:cden}

Column density of $^7$Be\,II and Ca\,II will be found using the spectra of 
  nova V5668 Sgr on day 58 (Molaro et al. 2016).
To this end we adopt that absorption lines form  
 in a plane slab on the background of a continuum source with the flux $F_c$
 at the inner boundary of the slab.
In fact continuum photons experience resonance scatter, not absorption. 
However, since the scattered photons are distributed in the broad range of 
  radial velocities ($\pm 10^3$\kms), the contribution of scattered photons in 
  the narrow line profile is negligible.
The scattering therefore can be treated as the absorption.
The absorbing medium is characterized by the gas temperature of 5000\,K 
 and microturbulent velocity $v_t$.
The turbulence includes also a possible radial velocity gradient in the shell.
Preliminary modelling shows that one need to introduce the contribution of 
  the unabsorbed (veiling) continuum in the wavelength range of the line profile in order 
  to describe a significant residual intensity of absorption lines. 
The fraction of the unabsorbed continuum is specified by the parameter $\eta$ 
   so that $\eta F_c$ is the flux avoiding line absorption and $(1-\eta)F_c$ 
   the flux passing through the absorbing gas.
The parameter $\eta$ can be considered as the fraction of "holes" in the shell 
  on the background of the continuum source. 
This is only an illustration of a possibility; the real geometry can be 
  more complicated and include large scale deviations from the spherical 
  symmetry typical for nova shells.
Another feature of the model is the need to use two Gaussian turbulent components 
  with microturbulent velocities $v_{t,1}$ and $v_{t,2}$ and corresponding 
  column densities $\phi_1 N$ and $(1 - \phi_1)N$, where $N$ is the total 
  average column density of $^7$Be\,II in the shell.
  
Modelling results for the doublet $^7$Be\,II 3130.42, 3131.07\,\AA\ are shown in 
 Fig. 2.
The model with $\eta = 0$ and $v_{t,1} = 12$\kms\ (model A in Table 1) fails to 
  reproduce both absorption lines of the doublet: the blue component is reproduced, 
  whereas the model for the red line has lower depth compared to the observed 
  profile. 
This reflects the absorption saturation in the presence of the 
  of the unabsorbed continuum.
A reasonable description is attained in the model B (Table 1) with $\eta = 0.6$. 
The presence of the second component with $v_{t,2} = 5$\kms\ is related to the narrow 
  width of lower part of the absorption profile. 
Column density of the model B is 5.5 times greater than that of the model A. 
After allowing for the holes this factor reduces down to 2.2.

The model B is used for the description of the Ca\,II 3933\,\AA\
 absorption in the same spectrum.
The found column density of Ca\,II is $6\cdot10^{11}$\,cm$^{-2}$.
This value is used below for the determination of the $^7$Be abundance. 
The ratio of column densities of Be\,II and Ca\,II on day 58 is 
  $N(^7\mbox{BeII})/N(\mbox{CaII}) =  36.6$, close to the value of 31.9 
  estimated by Molaro et al. (2016) from equivalent widths of 
 Ca\,II 3933\,\AA\ and $^7$Be\,II 3130\,\AA.

%=============================================================
\begin{table}[t]
	
	\vspace{6mm}
	\centering
	{{\em Table 1.} Parameters of models for the synthetic spectra of 
	the $^7$Be\,II doublet}
	\label{tab:spar} 
	
	\vspace{5mm}\begin{tabular}{l|c|c|c|c|c|c} 
		\hline
Модель & $\eta$   & $N~(10^{12}$\,cm$^{-2}$) &  $v_{t,1}$ (\kms)  & $v_{t,2}$ (\kms)  & $\phi_1$  \\
\hline
    A   &   0     &   4      &   12       &   5          &    1        \\
    B   &   0.6   &   22     &   12       &   5          &    0.55      \\
		\hline
		
	\end{tabular}
\end{table}
%===================================================================

\subsection{Be and Ca ionization and $^7$Be mass}

The ionization of Be and Ca in the shell of V5668 Sgr at $t= 58$\,d is considered 
   assuming that the shell of the radius $r = vt$ with the expansion velocity of  
   $v = 1175$\kms\ is irradiated by the diluted black-body radiation with the 
    temperature $T$ emitted by the photosphere of the radius $r_p$.
In this case the ionization is described by the modified Saha equation with the 
  dilution factor $W = 0.5[1 - (1-(r_p/r)^2)^{0.5}]$
\begin{equation}  
n_e\frac{n_{k+1}}{n_k} = W\left(\frac{T_e}{T}\right)^{1/2}S_{eq}(T)\,,
\end{equation}
where $n_e$ is the electron number density, $n_k$ is the ion concentration 
in $k$-th ionization state ($k = 1$ corresponds to neutrals), 
 $S_{eq}(T)$ is the r.h.s. of Saha equation at the thermodynamic equilibrium, 
 $T_e = 5000$\,K is the electron temperature in line with the kinetic temperature 
  used in the line profile model.
Throughout 100 days the bolometric luminosity of V5668 Sgr remained at the level 
  of  $\approx 2\cdot10^{38}$\ergs\ (Gehrz et al. 2018); this value is adopted below.
The photosphere temperature and dilution factor can be determined given the 
  photosphere  radius that is not known yet. 
This can be determined based on the analysis of the Ca ionization and Ca column density.   

The lower limit of the shell mass can be found using dust mass estimate in the 
   shell of V5668 Sgr,  $\sim 1.2\cdot10^{-7}$\msun\ (Gehrz et al. 2018).
The carbon in nova DQ Her 1934 shows excess of 1 dex compared to the 
  solar abundance (Mustel and Baranova 1966). 
Nova V5668 Sgr is similar to DQ Her in many respects, so one can expect similar 
   carbon overabundance.
If all the carbon is converted to the dust then the shell mass of V5668 Sgr 
  should be $\sim 10^{-6}$\msun.
This is lower limit since not all the carbon is converted to the dust.
Indeed this is indicated by the presence of the emission of C\,II 7231, 7236, 7237\,\AA\ 
 on day 114 (Harvey et al. 2018) at the stage when the dust mass was maximal 
  (Gehrz et al. 2018).
We adopt the shell mass for the fiducial model of $10^{-5}$\msun\ admitting 
  deviations by a factor of two in both sides.
  
The ionization fraction of Be\,II and Ca\,II is calculated for different values of 
 the photospheric radius.
Six ionization stages are included for calcium and four stages for beryllium.
The relative width of the shell is assumed to be $\delta = \Delta r/r = 0.1$ which  
  is consistent with the small velocity dispersion of the shell ($\Delta v/v < 0.1$) 
  according to beryllium lines.
The shell density is determined with the hole fraction $\eta = 0.6$ in 
  accord with the $^7$Be\,II doublet model.
The electron number density suggests the complete hydrogen ionization for the solar   
  composition; deviation from the solar composition does not affect result 
  significantly.  
  
The calculated ionization fraction of Be\,II and Ca\,II is shown in Fig. 3a for the 
 fiducial shell of $10^{-5}$\msun.
The different behavior for Be and Ca is related to relatively low ionization potentials 
  of Ca\,II and Ca ions with charges $> 1$.
Figure 3a implies that the ionization fraction of $^7$Be exceeds the 
  fraction of Ca\,II by at least 1\,dex which makes the equality of both 
    fractions in novae very unlikely.
It is reasonble to assume that Ca abundance in nova shell is solar since calcium 
 is not synthesized in novae flash.
Taking into account adopted shell mass and the found column density of Ca\,II  
  in the section \ref{sec:cden} we can find the photospheric radius for which 
  the fraction Ca\,II/Ca corresponds to solar abundance of Ca; it is   
  marked by dotted line in Fig. 3a.
The radius corresponds to the photospheric temperature  of $\approx 15000$\,K. 
The crossing of the dotted line with the curve $f(^7\mbox{Be})$ results in 
  the ionization fraction of $^7$Be\,II that being combined with the column density 
 of $^7$Be\,II gives us the mass of $^7$Be and accordingly mass fraction of 
 $^7$Be for the fiducial shell mass.
 
 The described method for the determination of the mass fraction of $^7$Be  
  is realized for the shell mass range of $(0.5-2)\cdot10^{-5}$\msun\ (Fig. 3b).
The $^7$Be abundance is referred to the moment of the thermonuclear flash.
The interval of X$(^7\mbox{Be})$, for which the column density of Ca\,II is 
  equal to the value found from the absorption line of 
   Ca\,II 3933 \AA\  is shown by the section between extreme masses.
In the considered mass range the $^7$Be mass fraction is confined in the 
  range of  $2.7\cdot10^{-6} - 2.8\cdot10^{-5}$.
For the fiducial shell mass of $10^{-5}$\msun\ the $^7$Be mass fraction is 
  equal to $8\cdot10^{-6}$. 
It is remarkable that the found abundance of X$(^7\mbox{Be})$ for the shell 
  mass range 
  of $(0.5-2)\cdot10^{-5}$\msun\ is consistent with theoretical predictions.

\section{478 keV gamma-ray line flux}

$^7$Be decay into $^7$Li via $e^{-}$-capture is accompanied by the emission of 
  gamma-quanta of 478 keV with the probability of 0.105 (Firestone et al. 1999).
Nova V5668 Sgr came in view of the spectrometer SPI on board of {\em INTEGRAL} 
  (Siegert et al. 2018) during 44 days after the thermonuclear flash 
  with the total exposure of $10^6$ s. 
The gamma-line is not detected with the $3\sigma$ uper limit of 
   $8.2\cdot10^{-5}$ ph.\,cm$^{-2}$\,s$^{-1}$.
   
This upper limit can be compared to the expected flux at 478 keV for the model 
  with shell mass of $10^{-5}$\msun\ and the $^7$Be abundance of 
  X$(^7\mbox{Be}) = 8\cdot10^{-6}$. 
The $^7$Be abundance is assumed homogeneous across the shell.
The shell presumbly forms as a result of the constant mass loss 
  rate with the constant velocity of $v = 1175$\kms\ from the radius 
  $r_0 =10^{11}$\,cm which value is not important. 
The shell density is determined by the wind kinetic luminosity 
 $L_w = (1/2)wv^3$, where $w = \dot{M}/v$ is the wind density parameter.
We consider two cases: $L_w = 10^{38}$\ergs\ and $L_w = 2\cdot10^{38}$\ergs.
These options suggest the duration of the wind flow 16 and 8 days respectively.
The model flux for the distance of 1.2 kpc (Gehrz et al. 2018) along with 
the upper limit is shown in Fig. 4. 
For the adopted model the predicted flux in 478 keV line is 1.5 dex lower 
 than the observational upper limit.

\section{Discussion and conclusions}

The goal of our paper is to explore a reason for the disparity between 
  theoretical predictions of the  $^7$Be abundance in novae shells and recent 
  estimates of the $^7$Be abundance in novae based on the detected 
  absorption lines of the doublet $^7$Be\,II 3130.42, 3131.07\,\AA.
Our study of the thermonuclear burning in novae based on the two-zone model 
  shows that the $^7$Be abundance in nova shells cannot exceed $3\cdot10^{-5}$.
This limit is three times lower than the observational estimate for the 
 nova V5668 Sgr on day 58 after the light maximum (Molaro et al. 2016). 
Since the latter estimate is obtained assuming similar ionization fraction 
  of Be\,II and Ca\,II we find it necessary to reassess the $^7$Be abundance 
  estimate using the same spectrum. 
The new estimate taking into account ionization of Be and Ca shows that 
  the ratio Be\,II/Be at least by 1\,dex greater than the ratio Ca\,II/Ca 
  and therefore the assumption for the equality of these ratios is unjusified.
The primary conclusion of our study is that the  $^7$Be abundance
  for the mass range of nova shell of $(0.5-2)\times10^{-5}$\msun\ is consistent with the theoretical predictions.
 
For the fiducial shell mass of $10^{-5}$\msun\ the flux of 478 keV gamma-quanta 
   from $^7$Be decay is by 1.5\,dex lower than the upper limit suggested by 
  {\em INTEGRAL} observation of nova V5668 Sgr.
Currently there is no contradiction between the observational estimate of 
  the $^7$Be abundance in nova V5668 Sgr and both theoretical predictions and 
  478 keV gamma-line observation.
   
It is noteworthy that our computation of Be and Ca ionization is based on the 
  approximation of the ionized radiation by the black-body spectrum.
More refined description of the ionizing radiation is needed for the 
  reliable estimate of $^7$Be abundance in novae from absorption lines. 
The adequite models of novae and growing number of explored events hopefully 
  will permit us to infer the relibale estimate of novae contribution in the 
  galactic synthesis of $^7$Li.
  
Here we bound ourselves with a crude estimate.
Assuming the shell mass of $10^{-5}$\msun\ to be typical for classical novae
  and using the estimated abundance X$(^7\mbox{Li}) = 8\cdot10^{-6}$ one gets 
  the typical mass of $^7$Li produced per one nova of $8\cdot10^{-11}$\msun.
With the rate of classical novae in the Galaxy of  $\approx50$ yr$^{-1}$ 
  (Shafter et al. 2017) during the latest $10^{10}$ yr novae should synthesise 
   $\approx40$\msun\ of $^7$Li.
For the mass of barion matter in the Galaxy of $6.08\cdot10^{10}$\msun\ 
  (Licquia and Newman 2015) we find the present day galactic $^7$Li abundance 
  due to novae $A(^7\mbox{Li}) = 12 - \log{(N(\mbox{Li})/N(\mbox{H})}) \approx 2.15$ 
  which is only 8\% of the meteoritic $^7$Li abundance $A(^7\mbox{Li}) = 3.26$ 
  (Grevesse et al. 2010). 
Allowing for the uncertainty of novae shell masses one could admit that the average 
  shell mass is higher, e.g., $2\cdot10^{-5}$\msun.
In this case using X$(^7\mbox{Li}) = 2.8\cdot10^{-5}$ we get 
  $^7$Li mass per one nova of $5.6\cdot10^{-10}$\msun, which suggests 
  that about 50\% of galactic $^7$Li could be synthesised by novae.
Currently the issue of the nova role in the galactic $^7$Li thus is highly 
 uncertain with the possible contribution between 
  $\lesssim 10$\% and $\sim 100$\%.

%***************************************************************

\pagebreak   
%****************************************************************

\pagebreak   
%****************************************************************

%xxxxxxxxxxxxxxxxxxxxxxxxxxxxxxxxxxxxxxxxxxxxxxxxxxxxxxxxxxxxxx
\clearpage
%xxxxxxxxxxxxxxxxxxxxxxxxxxxxxxxxxxxxxxxxxxxxxxxxxxxxxxxxxxxxxx
\begin{figure}[h]
	\epsfxsize=19cm
	\hspace{-2cm}\epsffile{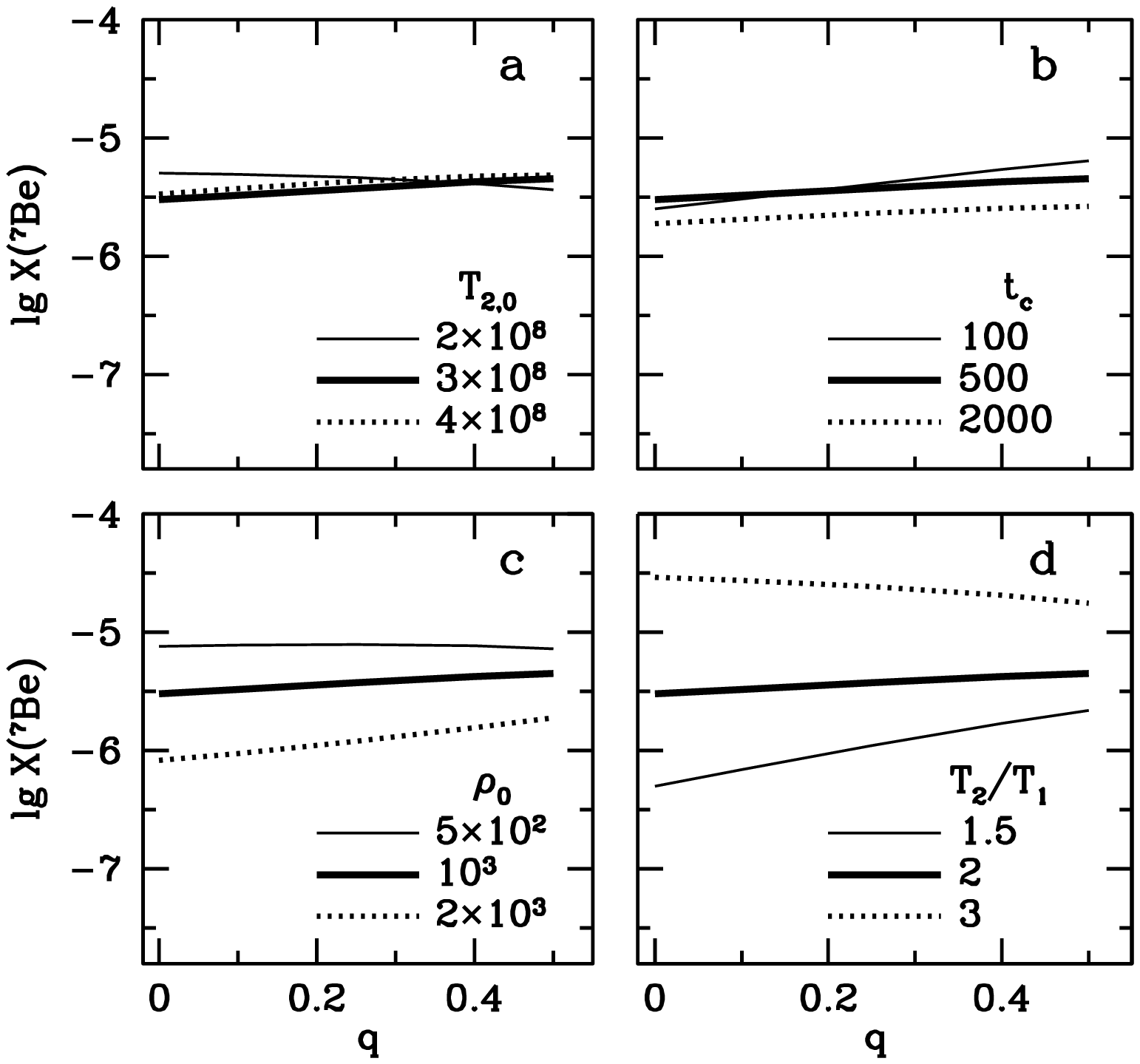}
	\caption{\rm  
	$^7$Be abundance in two-zone model versus fraction of the white dwarf 
	matter in the hydrogen envelope. Panel {\em a} shows an effect of 
	the initial temperature in the central zone; panel {\em b} shows an effect 
	of the mixing time (in seconds); panel {\em c} shows an effect of the 
	initial density (\gcm) in the central zone; panel {\em d} shows an effect 
	of the ratio of the temperature in the central zone to outer zone.
	}
\end{figure}
%========================================================
\clearpage
%xxxxxxxxxxxxxxxxxxxxxxxxxxxxxxxxxxxxxxxxxxxxxxxxxxxxxxxxxxxxxx
\begin{figure}[h]
	\epsfxsize=19cm
	\hspace{-2cm}\epsffile{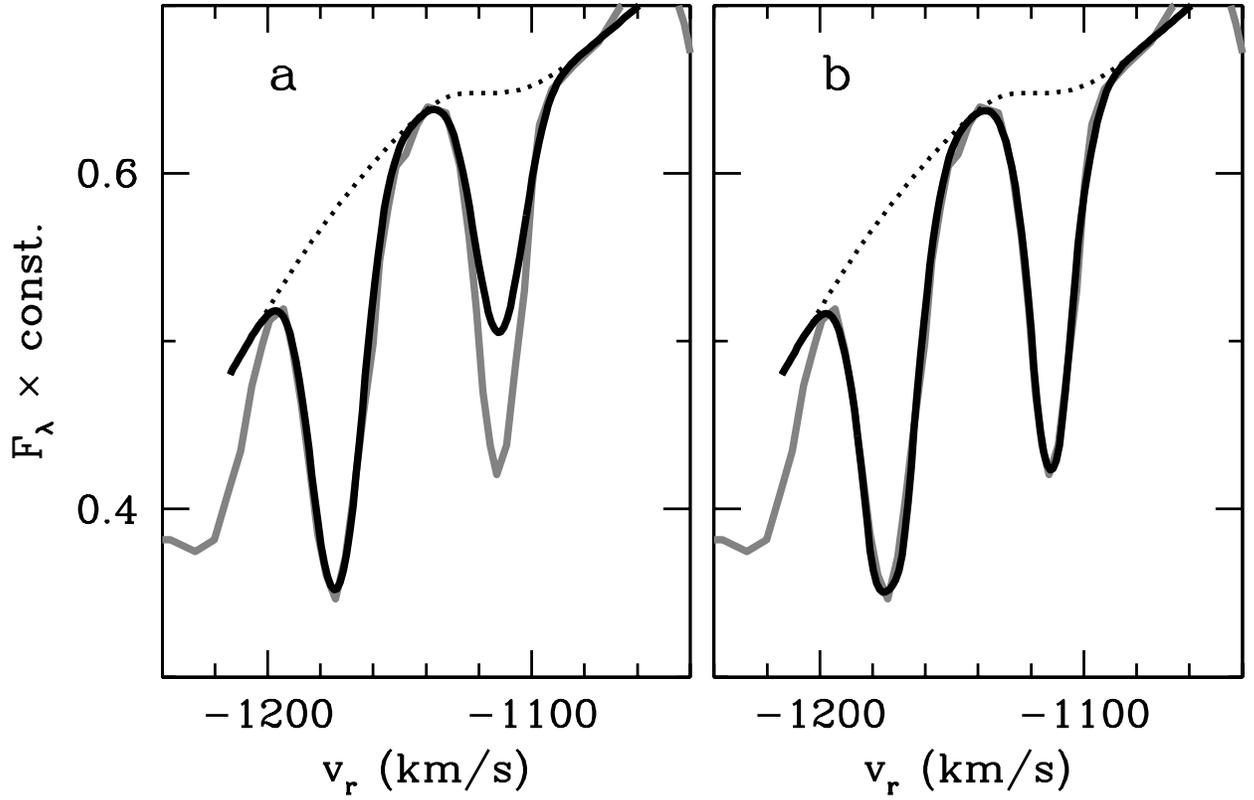}
	\caption{\rm 
	The model and observed spectrum of nova V5668 Sgr on day 58 
	({\em grey}) line in the region of the $^7$Be 3130.42, 3131.07\,\AA\ 
	doublet. {\em Dotted} line shows adopted quasicontinuum. 
	Panel {\em a} shows the case without holes in the shell (model A), 
	whereas panel {\em b} shows the case with holes against the continuum 
	background.
	}
\end{figure}
%========================================================

\clearpage
%xxxxxxxxxxxxxxxxxxxxxxxxxxxxxxxxxxxxxxxxxxxxxxxxxxxxxxxxxxxxxx
\begin{figure}[h]
	\epsfxsize=19cm
	\hspace{-2cm}\epsffile{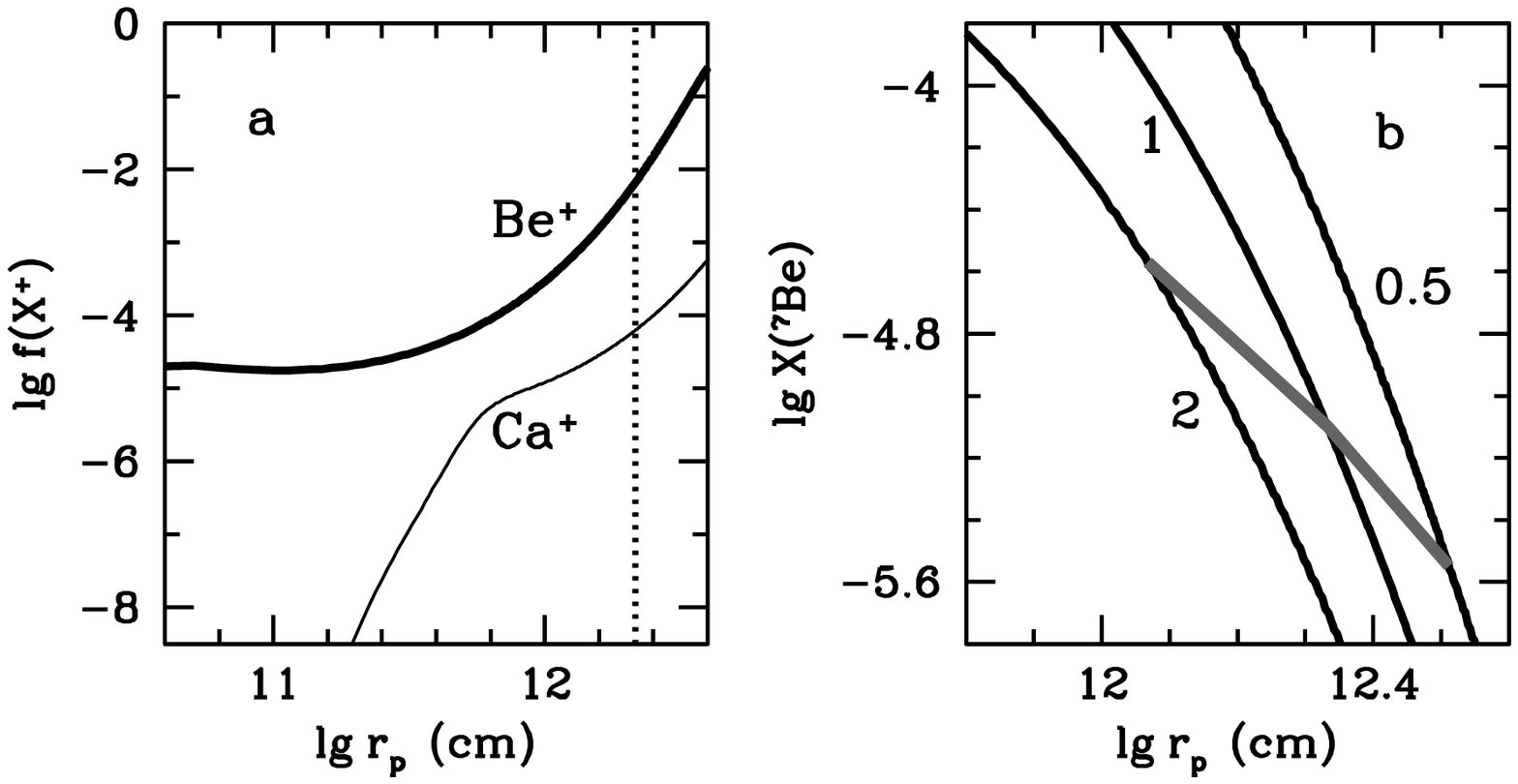}
	\caption{\rm  
	Ionization fraction of Be\,II and Ca\,II and the $^7$Be abundance.
	Panel {\em a} shows the ionization fraction of Be\,II and Ca\,II 
	in the nova shell depending on the photospheric radius adopting the 
	luminosity of $2\cdot10^{38}$\ergs, shell mass of $10^{-5}$\msun, and 
	the expansion velocity of 1175\kms. Panel {\em b} shows the mass $^7$Be 
	fraction for shell mass of 0.5, 1, and 2 in units of  $10^{-5}$\msun\ 
	versus the photospheric radius. {\it Grey} line is the loci of 
	X$(^7\mbox{Be})$ 
	values for the photospheric radius and shell mass that satisfy the condition of the solar Ca abundance.
	}
\end{figure}
%========================================================

\clearpage
%xxxxxxxxxxxxxxxxxxxxxxxxxxxxxxxxxxxxxxxxxxxxxxxxxxxxxxxxxxxxxx
\begin{figure}[h]
	\epsfxsize=19cm
	\hspace{-2cm}\epsffile{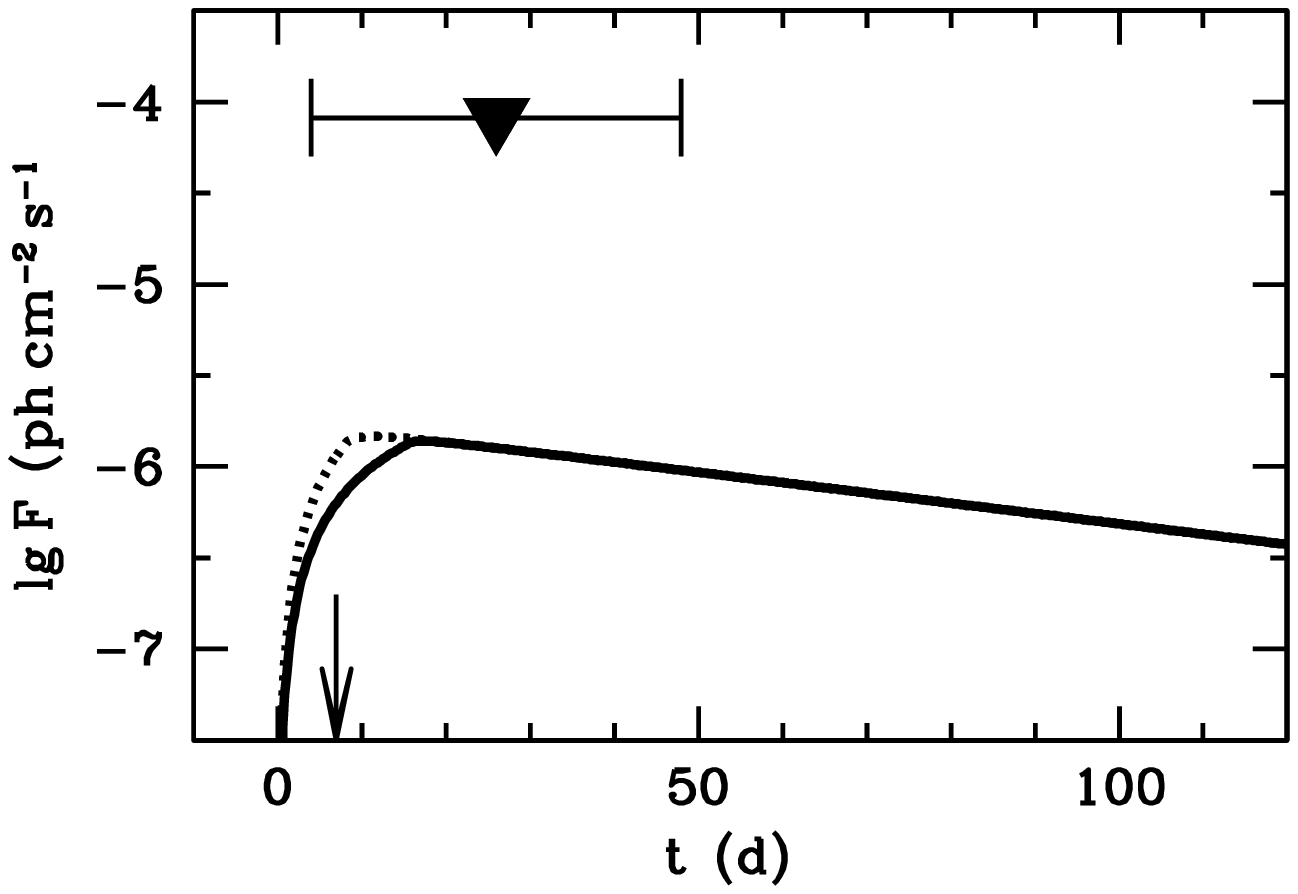}
	\caption{\rm  
	Flux of gamma-quanta in 478 keV line from the nova with parameters of V5668 Sgr
	(cf. text) and the distance of 1.2 kpc. The upper part of the figure shows 
	the upper limit according to {\em INTEGRAL} observation (Siegert et al. 2018) 
	with the indication of the observation epoch. The time is counted from the
	thermonuclear flash, i.e. 7 days prior to the light maximum (shown by arrow).
	Two model cases correspond to the wind kinetic luminosity $L_w = 10^{38}$\ergs\ 
	({\em solid} line) and $2\cdot10^{38}$\ergs.
	}
\end{figure}
%========================================================


\begin{thebibliography}
%=========================================== 
\bibitem{}
  Angulo C., Arnould M., Rayet M. et al., Nuclear Physics A {\bf 656}, 3 (1999)
\bibitem{}
  Boffin H.M.J., Paulus G., Arnould M., Mowlavi N., 
  Astron. Astrophys. {\bf279}, 173 (1993)
\bibitem{}
  Cameron A.G.W.  Astrophys. J. {\bf121}, 144 (1955)  
\bibitem{}
  Cyburt R.H., Amthor A.M., Ferguson R., Astrophys. J. Suppl. {\bf189}, 240 (2010)  
\bibitem{}  
  Darnley M.J., Bode M.F., Kerins E. et al., Mon. Not. R. Astron. Soc. 
  {\bf 369}, 257 (2006)
\bibitem{}
  Denissenkov P.A., Truran J.W., Pignatari M. et al., 
  Mon. Not. R. Astron. Soc. {\bf442}, 2058 (2014) 
\bibitem{} 
  Firestone R.B., Shirley V.S., Baglin C.M. et al.,
  {\it Table of Isotopes} (John Wiley and Sons, New York, 1999)  
 \bibitem{}  
  Gehrz R. D., Evans A., Woodward C. E.  et al.Astrophys. J. {\bf 858}, 78 (2018)
\bibitem{} 
  Grevesse N., Asplund M., Sauval A.J., Scott P., Astrophys. Space Sci {\bf328},
  179 (2010)
\bibitem{}
  Hernanz M., Jos{\'e} J., Coc A., Isern J., Astrophys. J. {\bf465}, L27 (1996) 
\bibitem{}
  Izzo L., Molaro P., Bonifacio P. et al.,  Mon. Not. R. Astron. Soc. 
  {\bf478}, 1601 (2018) 
 \bibitem{}
  Jos{\'e} J., Hernanz M., J. Phys. G Nucl. Phys. {\bf34}, 431 (2007)
\bibitem{}
  Kudryashov A.D., INASAN Science Reports {\bf 3}, 205 (2019)
\bibitem{}
  Kudryashov A.D., Chugai N.N., Tutukov A.V., Astron. Rep. {\bf44}, 170 (2000)
\bibitem{}
  Licquia T.C., Newman J.A., Astrophys. J. {\bf806}, 96 (2015)
\bibitem{}
  Molaro P., Izzo L., Mason E., Bonifacio P., Della Valle M.,
  Mon. Not. R. Astron. Soc.  {\bf463}, L117 (2016)
 \bibitem{} 
  Mustel E.R., Baranova L.I., Soviet Astron. {\bf 10}, 388 (1966) 
\bibitem{}
  Ritossa C., Garcia-Berro E., Iben I.Jr., Astrophys. J. {\bf460}, 489 (1996)
\bibitem{}
  Selvelli P., Molaro P., Izzo L., Mon. Not. R. Astron. Soc. 
  {\bf481}, 2261 (2018)
\bibitem{}
  Seach J., Cent. Bur. Electron. Telegrams, 4080 (2015)
\bibitem{}
  Shafter A.W., Astrophys. J. {\bf 834}, 196 (2017)
\bibitem{}
  Siegert T., Coc A., L. Delgado et al., Astron. Astrophys. {\bf625}, 1075 (2018)
\bibitem{} 
  Starrfield S., Bose M., Iliadis C. et al., arXive:1910.00575v1 (2019)
\bibitem{} 
  Starrfield S., Truran J.W., Sparks W.M., Astrophys. J. {\bf226}, 186 (1978)
\bibitem{} 
  Tajitsu A., Sadakane K., Naito H., Astrophys. J. {\bf818}, 191 (2016)
\bibitem{}
  Tajitsu A., Sadakane K., Naito H. et al., Nature {\bf518}, 381 (2015)

%=====================================================

\end{thebibliography}
\end{document}